\documentclass[twocolumn,10pt]{asme2ej}

\usepackage{epsfig} 
\usepackage[english]{babel}
\usepackage{algpseudocode}
\usepackage{graphicx}
\usepackage{caption}
\usepackage{subcaption}
\usepackage[toc,page]{appendix}

\usepackage{titlesec}
\titlespacing\section{0pt}{10pt plus 4pt minus 2pt}{4pt plus 2pt minus 2pt}
\titlespacing\subsection{0pt}{10pt plus 4pt minus 2pt}{4pt plus 2pt minus 2pt}

\usepackage[centertags,reqno]{amsmath}
\usepackage{amssymb,bm,amsfonts}
\usepackage{mathtools}
\usepackage[thmmarks,amsmath]{ntheorem}

\newtheorem{remark}{Remark}

\newtheorem{prop}{Proposition}

\usepackage{enumerate}

\usepackage{amsmath,array}
\usepackage{mathtools, nccmath}

\usepackage{mathtools} 
\usepackage{graphicx}
\usepackage{cite}
\usepackage[section]{algorithm}
\usepackage[]{fancyhdr}

\fancypagestyle{firstpage}{%
  \lhead{**Working Draft, not yet peer-reviewed**}
  \rhead{}
}

\makeatletter
\newcommand{\distas}[1]{\mathbin{\overset{#1}{\kern\z@\sim}}}
\newsavebox{\mybox}\newsavebox{\mysim}
\newcommand{\distras}[1]{
  \savebox{\mybox}{\hbox{\kern3pt$\scriptstyle#1$\kern3pt}}
 \savebox{\mysim}{\hbox{$\sim$}}
  \mathbin{\overset{#1}{\kern\z@\resizebox{\wd\mybox}{\ht\mysim}{$\sim$}}}
}
\makeatother

\renewcommand{\u}{\mathbf{u}}

\newcommand{\R}{\mathbb{R}}

\renewcommand{\x}{\mathbf{x}}

\renewcommand{\r}{\mathbf{r}}

\renewcommand{\v}{\mathbf{v}}
\newcommand{\w}{\mathbf{w}}

\newcommand{\y}{\mathbf{y}}

\title{GLHAD: A Group Lasso-based Hybrid Attack Detection and Localization Framework for Multistage Manufacturing Systems}

\author{Ahmad Kokhahi
    \affiliation{
	PhD Student\\ Department of Industrial Engineering\\ Clemson University\\
	  Email: akokhah@clemson.edu
    }	
}

\author{Dan Li 
    \affiliation{Assistant Professor\\ Department of Industrial Engineering\\ Clemson University\\
        Email: dli4@clemson.edu
    }
}

\begin{document}
\setlength\abovedisplayskip{5pt}
\setlength\belowdisplayskip{5pt}

\maketitle    
 \thispagestyle{firstpage}
\begin{abstract}
{\it 
As Industry 4.0 and digitalization continue to advance, the reliance on information technology increases, making the world more vulnerable to cyber-attacks, especially cyber-physical attacks that can manipulate physical systems and compromise operational data integrity. Detecting cyber-attacks in multistage manufacturing systems (MMS) is crucial due to the growing sophistication of attacks and the complexity of MMS. Attacks can propagate throughout the system, affecting subsequent stages and making detection more challenging than in single-stage systems. Localization is also critical due to the complex interactions in MMS. To address these challenges, a group lasso regression-based framework is proposed to detect and localize attacks in MMS. The proposed algorithm outperforms traditional hypothesis testing-based methods in expected detection delay and localization accuracy, as demonstrated in a simple linear multistage manufacturing system.
}
\end{abstract}


\section{Introduction}\label{sec1}

Due to the advancement in automation and the industrial internet-of-things (IIoT), concerns around the cybersecurity of manufacturers have grown considerably in recent years\cite{mahoney2017cybersecurity}. After the well-known Stuxnet \cite{langner2011stuxnet} targeting Iran's nuclear program, cyberattacks aiming at disrupting manufacturing operations have surged. In 2022, the manufacturing industry became the top-target of cyberattacks in all operational technology (OT)-related industries. WannaCry \cite{wu2019intrusion} ransomware affecting several companies such as Taiwan Semiconductor Manufacturing  Company (TSMC), and an attack against the German still mill company in 2014\cite{lee2014german} have caused enormous problems for the countries and industries in recent years. The above attacks are mostly cyber-physical attacks, where the attack intrudes into the system from the cyber network but aims to disrupt the physical process. Examples of cyber-physical attacks include eavesdropping, denial-of-service (DoS) attacks, stealthy deception attacks, jamming attacks, compromised-key attacks, and man-in-the-middle (MITM) attacks\cite{singh2020review}. In the above attacks, both the stealthy deception attacks and MITM involved malicious data manipulations, which affect the integrity of the system and potentially compromise product quality and system safety in manufacturing. Therefore, in this paper, we focus on detecting and localizing these data integrity attacks on the sensor data.

Multistage manufacturing systems are very important in the manufacturing industry\cite{liu2021cyber}. These systems, which consist of multiple components, stations, or stages, can be modeled as a series of interconnected elements working together to produce the final product\cite{liu2021cyber,shi2006stream}. The digital connectivity between components and devices in MMSs as well as the standard data communication protocols used in Manufacturing Execution Systems (MES) makes the modern manufacturing systems vulnerable to cyber-attacks. Besides, due to the interconnectivity between stages in an MMS, a single point of failure can quickly spread throughout the system and lead to quality issues in the final product, highlighting the importance of robust cyberattack detection and localization measures. For example, in a car assembly process, an attack on the machining stage will change the dimension of the part under assembling, causing the final product not to serve its purpose and consequently yielding quality issues in the final product. Therefore, it is important to detect cyberattacks in the early stage and localize the stage under attack. 

 Most of the existing studies on cybersecurity of multistage manufacturing systems consider MMS in the context of additive manufacturing \cite{al2022securing,shi2022lstm,zeltmann2016manufacturing,liu2020online}. However, such methods are specific to additive manufacturing and can not be applied to generic MMS such as assembly processes. For example, in \cite{al2022securing}, the printing of each layer is considered as a stage, and a layer-wise alteration detection method is developed based on the image analysis after each layer is printed. Such methods are generalizable to other layer-by-layer manufacturing processes. However, there lacks a fundamental study that analyses the mechanism of how cyberattacks propagate to other stages and how we can leverage such propagation for attack localization. Therefore, we aim to analyse the attack propagation in a generic multistage process and develop a cyberattack detection algorithm that is generalizable to multiple MMS applications.


This paper considers a generalized multi-stage manufacturing system, where each stage sequentially processes the product. Each stage consists of a set of sensors, which take measurements of the product, and a controller that calculates the control output based on the sensor measurements from the previous stage to ensure the output sensor measurements of the stage are at the desired level. Because of this control mechanism, the impact of false data injection attacks in a specific stage can propagate to later stages. That is, a false data injection in the previous stage may not cause a physical impact on the attacked stage but on later stages. Moreover, when the attacker has some knowledge about the system, the attack can be designed to be undetectable at that stage\cite{liu2011false}. The above factors pose significant challenges in detecting and localizing the attack in MMS, which we aim to address in this paper. The contributions of the paper can be summarized as follows:
\begin{enumerate}
    \item We characterize a generic multi-stage manufacturing process model using Kalman filter(KF) and stochastic state space model and perform theoretical analysis to extract the features uniquely related to the location of the attack.
    \item We design a hybrid detection framework that integrates the benefits of signature- and anomaly-based detection techniques that fulfills the localization function without relying on attack data for training. This is achieved based on the theoretical analysis, which extracts the feature based on the domain knowledge of the system dynamics.
    \item We designed a Group regularization-based framework for simultaneous attack detection and localization. Unlike most existing methods that detect and localize in a two-phase manner, the group lasso-based framework enables us to identify the occurrence and location of potential false data injection attacks in real time. With such information, further investigation and treatments can be triggered to minimize the attack's impact on the system and its users.  
\end{enumerate}

\section{Literature Review}\label{li}

Cyberattacks in manufacturing systems lie in the domain of cyber-physical attacks. While there are many security approaches regarding cyber-physical systems (CPSs), including vulnerability analysis \cite{northern2021vercasm,zhang2022modeling,pan2020modeling}, secure IoT network architecture design \cite{pivoto2021cyber,patan2020smart}, intrusion detection \cite{thakur2021intrusion,althobaiti2021intelligent}, and cyberattack-resilient state estimation and control \cite{kazemi2021finite,ding2020secure,zhao2022anti}, we focus on the process-based cyber-attack detection as it is the subject of this paper.


The literature in the area of data-driven cyberattack detection in manufacturing systems can be categorized as ($i$) signature-based methods and ($ii$) anomaly-based methods. The signature-based methods work based on the known attacks and raise the alarm when a pattern matches a known attack \cite{wu2019intrusion,liao2013intrusion,yaacoub2020cyber}. In other words, it works like supervised machine learning algorithms; they are trained based on normal and under-attack data. Then, the trained algorithm is deployed on real-world data for detection of attacks\cite{panigrahi2022intrusion,kwon2022advanced,song2020layered,wu2016detecting}
Song et al. propose a real-time attack detection system using a convolutional neural network (CNN) in Cyber-manufacturing systems (CMS) for detecting defects \cite{song2019physical}. Wu et al. deploy machine learning algorithms for detecting cyber-physical attacks in cyber manufacturing systems. They
use simulated 3D printing and CNC machining malicious attacks\cite{wu2019detecting}. Wu et al. try to detect malicious infill defects in 3D printing with the help of image classification. They first extract features from the images and
then apply classification algorithms, namely naive Bayes classifier and J48 Decision Tree\cite{wu2016detecting}.

Anomaly detection methods try to extract patterns for the system's expected behavior and raise the alarms when recognizing any significant deviation from the normal pattern. In other words, these methods work like unsupervised learning algorithms\cite{kwon2022advanced,bhardwaj2020capturing,qian2020cyber,abokifa2019real}. Qian et al. propose a scheme for detecting cyber attacks in the cyber and physical stages of Supervisory Control and Data Acquisition (SCADA) systems by using a Nonparallel hyperplane-based fuzzy classifier presented in the paper\cite{qian2020cyber}. Kwon et al. propose a hybrid of anomaly and signature detection algorithms for detecting cyber attacks in  physical systems. They use normal data in training datasets to find the threshold  and then apply the trained model to the test dataset to see the model's performance\cite{kwon2022advanced}.


The proposed method in this paper is a hybrid of signature and anomaly detection techniques. The signatures corresponding to each stage in the system characterize the correlation between sensor measurements of all stages and the injected data. However, the signatures are not extracted as data-driven as the supervised learning-based algorithms mentioned above. Still, they are derived from the system dynamics based on domain knowledge. Therefore, the proposed method is also considered an anomaly-based detection algorithm that only relies on the normal data rather than the attack data. Thus, the proposed method is more realistic in MMS applications as there is usually insufficient attack data.

\begin{figure*}[h]
\centering
    \includegraphics[width=0.8\textwidth]{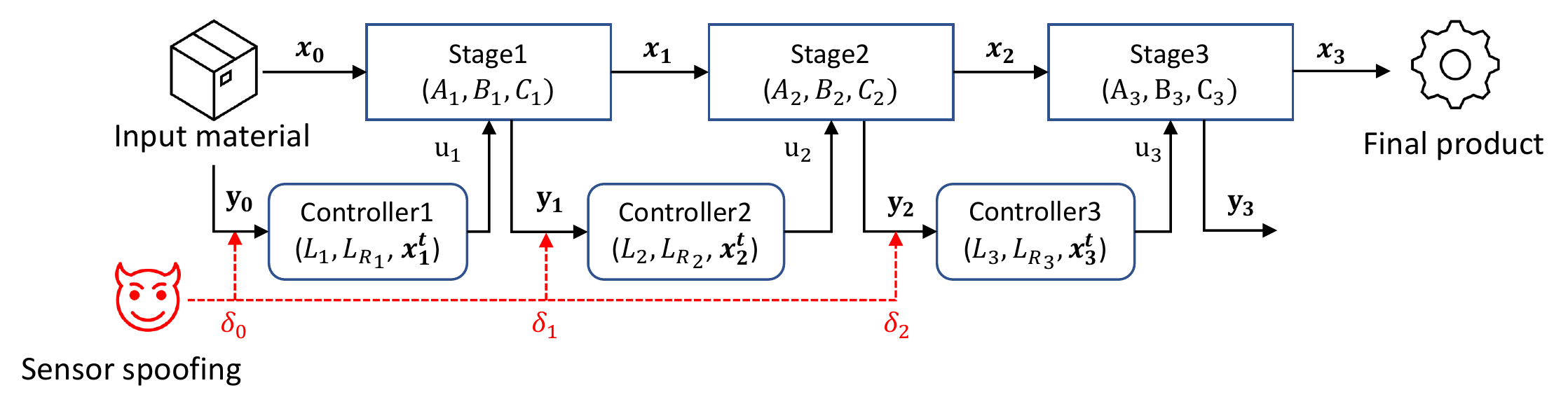}
\caption{System representation: attack injects false data to sensor }
\label{figure:sys_rep} 
\end{figure*}
\section{System Representation}\label{sec:3}
As discussed in Section \ref{sec1}, MMSs consist of multiple stages in which the output of stage $i$ is the input to stage $i+1$. The representation of the system can be seen in Fig.~\ref{figure:sys_rep}. At each stage, the control action is taken based on the measurements of the input state to control the states to the reference values (setpoints). In practice, measurements before and after a processing stage maybe taken at the same station. However, we generalize the measurements to be taken after each stage, which means the sensor measurements is obtained from the output state at each stage. The attack we are considering in this paper is false data injection attack, where fake sensor measurements are sent to the controllers.  
\subsection{State-Space Model}
The state-space representation has been widely used in the literature to characterize multi-stage processes.\cite{urbina2016limiting,li2020detection,mo2013detecting,van2015sequential}. Hence, we use a stochastic state-space model to represent an $K$-stage MMS. Let $\x_k$ denote the state variable of the product at stage $k$ such that $\x_k \in \R^{m_k}$, where $m_k$ is the number of state variables at stage $k$.  Let $\y_k$ denote the sensor measurements at stage $k$ such that $\y_k \in \R^{n_k}$, where $n_k$ is the number of sensors at stage $k$. Let $\u_k$ denote the control actions at time $k$ such that $\u_k \in \R^{p_k}$, where $p_k$ is the dimension of control action at stage $k$. $\w_k$ and $\v_k$ are the process noise and measurement noise terms at stage $k$, which are independent of all the other variables and are assumed to be not affected by any system anomaly. Both $\w_k$ and $\v_k$ follow multivariate normal distributions with zero mean, i.e., $\w_k \sim K(0,W_k)$ and $\v_k \sim K(0,V_k)$, where $W \in \R^{m_k\times m_k}$ and $V \in \R^{n_k \times n_k}$ are the covariance matrices. The state-transition function and the measurement function are:
\begin{gather}
     \x_k = A_k\x_{k-1} + B_k\u_k + \w_k,\label{eq:1}\\
     \y_k = C_k\x_k + \v_k.\label{eq:2}
\end{gather}
In the above equations, $A_k\in\R^{m_k\times m_{k-1}}$, $B_k\in\R^{m_k \times p_k}$, and $C_k\in\R^{n_k \times m_k}$ are the system matrix, input matrix, and output matrix, respectively. We assume the matrices $A_k, B_k, C_k, V_k, W_k$ for each stage $k$ are known. In \eqref{eq:1}, $A_k\x_{k-1}$ is the action taken on the input product (state variable) at stage $k$. Also, $B_k\u_k$ is the action taken at stage $k$ by the controller to make sure that the state variable is the desired (reference) state variable at stage $k$.

\subsection{Controller Model}
We consider a linear controller, where the control action at stage $k$ is calculated as follows: 
 \begin{gather}\label{eq:controller}
    \u_k = L_k\hat{\x}_{k-1|k-1} + L_{R_k}{\x_k^r}.
\end{gather} 
In the equation above, $\hat{\x}_{k-1|k-1}$ is the state estimation of the product from the previous stage, $k-1$, and $\x_k^r$ is the reference value (or in general, the control setting parameters) of the state at stage $k$. The linear controller calculates the control action as a linear combination of the estimated state and the reference value. The matrices $L_k$ and $L_{R_k}$ are the linear coefficients of the estimated state after the previous stage and the reference values, respectively.

As a typical example, a Linear quadratic Gaussian (LQG) Controller follows the above formulation and can be used to calculate control action based on the state estimation $(\hat{\x}_{k|k})$ and the reference state of each stage $(\x_k^r$). The controller is calculated based on the minimization of 
$ J = \mathbb{E} [ 
 (\x_K-\x^r_K)^TF(\x_K-\x^r_K) + \Sigma_{k=0}^{K-2} (\x_k-\x^r_k)^TU(\x_k-\x^r_k) + \u_k^TZ\u_k],$ where $Z$ and $U$ are positive-semi definite matrices defining the cost. 
In \eqref{eq:controller}, the controller regulator is: $
    L_k = {(B_k^TS_{k+1}B_k + Z)}^{-1}B_kS_{k+1}A_k
$. $S_k$ is calculated by the following matrix Riccati difference equation that runs backward in time: $
    S_k = A_k^T(S_{k+1} - S_{k+1}B_k{(B_k^TS_{k+1}B_k + Z)}^{-1}B_kS_{k+1})A_k + U  , S_K=F
$.

Coupled with the state-space model, the Kalman filter (KF) is an optimal state estimator for the stochastic linear state-space model \cite{mo2009secure}. Therefore, we use a KF  to estimate the system state. System state estimation is needed for the computation of the controller. The Kalman gain $(K_k$) is derived based on the KF formulations of discrete time. 
KF formulas are given in the following :
\begin{gather}
    \hat{\x}_{k|k-1} = A_k \hat{\x}_{k-1|k-1} + B_k\u_k\\
      \hat{\x}_{k|k}=
       \hat{\x}_{k|k-1} + K_k {\hat{\y}_k}\\
      {\hat{\y}_k}=
       \y_k - C_k \hat{\x}_{k|k-1}
       \label{eq:5}
\end{gather}
Where $\hat{\x}_{k|k-1}$ denotes the predicted system state given at time $k-1$, $\hat{\x}_{k|k}$ denotes the updated state estimation given the measurement at time $k$, $\y_k$, and ${\tilde{y}_k}$ denotes the residual at time $k$, where ${\tilde{y}_k}$ is the difference
between predicted and actual measurements, as shown in \eqref{eq:5}.

\subsection{False data injection}

In this paper, we consider the false data injection (FDI) attacks on one of the stages. FDI attacks are implemented by maliciously manipulating sensor measurements by either spoofing the data or directly sending fake sensor data to the controller \cite{li2019feasibility,wang2019review,jorjani2020graph}. In this paper, a false data injection attack at stage $k$ can be modeled as:
\begin{gather}\label{eq:attack_on_stage}
\y_k^{a} = \y_k + {\delta_k},
\end{gather}
where ${\delta_k}$ represents the bias introduced to the sensor measurements by FDI at stage $k$. ${y_k^{a}}$ is the vector of the under-attack sensor measurements at stage $k$.


\section{Methodology}
Based on the system model described in section \ref{sec:3}, we propose the GLHAD framework for attack detection and localization in MMS. We first perform theoretical analysis to build the mathematical model, based on which we extract the data features characterizing the impact of false data injection attacks on the sensor data. We then use the characterization to extract signatures from state estimation residuals that indicate the occurrence of the attack in the system and location in terms of the under-attack stage in the system. The formulation is performed as the Group lasso regression (GLR) model described in \ref{sub:4.2} and The GLHAD framework, which monitors the GLR coefficients based on in-stage $T^{2}$ tests.

 \subsection{Theoretical Analysis}\label{sub:4.1}
 To facilitate our analysis, we introduce the\textit{ augmented state variable} $ \tilde{\x}$ comprised of the input state and the reference values.
  \begin{gather}
  \tilde{\x}= [\x_0^T \quad ... \quad (\x_K^r)^T]^T,\label{eq:15}
 \end{gather}
 where $\x_0$ represents the input of the system, also ${x_k^r}$ represents the reference state variable at stage $k$. 
Notice the state variables $\x=[\x_0^T, \x_1^T\quad ... \quad \x_N^T]^T$ tracks the product state at each stage, and with the control actions at each stage calculated based on the measurements after the previous stage, the $\x_i$'s are cross-correlated. On the other hand, the augmented state variables are the independent external inputs to the MMS at each stage under our system setting. In this context, the sensor measurements,  $\y=[ {\y_{0}}^T, \y_1^T \quad ... \quad {\y_K}^T]^T$, with $\y_k$ representing the sensor measurements at stage $k $, for $ k \in \{0, ..., K\} $, can be represented as a linear function of $\tilde{\x}$ as follows:
 \begin{gather}\label{eq:12}
     \y=H \tilde{\x}+\epsilon,
 \end{gather}
where $H$ is the augmented measurement matrix. Specifically, $H$ characterizes the relationship between the augmented state variables and the sensor outputs. $H$ is comprised of $(K+1)\times(K+1)$ submatrices $h_{ij}$, characterizing the relationship between the sensor measurements in stage $i-1$ and the augmented state in stage $j-1$. We develop Proposition \ref{thm:1} to define matrix $H$.

\begin{prop}\label{thm:1}
For an MMS described in section \ref{sec:3}, the relationship between the sensor's measurements and state variables, in other words, Matrix $H$ in \eqref{eq:12} is represented as: 
    \begin{gather}
    \y = H\tilde{\x}\\H=[h_{ij}]_{(K+1)\times(K+1)}
  \end{gather}
  For $j \geq2$, we have :
    \begin{align*}
  h_{ij}=
  \left\{
\begin{array}{lcl}
 {0}, & {i=1,..,j-1}\\
{C_{i-1}B_{i-1}L_{R_{i-1}}},& {i=j}\\
{C}_{i}\prod_{b=j}^{i}({A}_{b}+{B}_{b}{L}_{b})B_{1}L_{R_{1}}, &  {i=j+1,...,K}
\end{array}\right\}
 \end{align*}
For $j=1$, we have :
 \begin{align*}
  h_{i1}=
  \left\{
\begin{array}{lcl}
 {C_{0}}, & {i=1}\\
{C_{1}[A_{1} + B_{1}L_{1}K_{0}C_{0}]},& {i=2}\\
\end{array}\right\}
 \end{align*}
 For $i\geq3$, we have :
 \begin{align*}
h_{i1} = &C_{i-1}[\prod_{k=1}^{i-2}(A_{i-k} + B_{i-k}L_{i-k})(A_{1} + B_{1}L_{1}K_{0}C_{0}) - ...\\  &  \sum_{j=1}^{i-3} [\prod_{k=1}^{j}(A_{i-k} + B_{i-k}L_{i-k})(B_{i-(j+1)}L_{i-(j+1)})...\\
&\prod_{c=j+2}^{i-1}(I - K_{i-c}C_{i-c})A_{i-c}(I - K_{0}C_{0})] - ...\\&(B_{i-1}L_{i-1})\prod_{m=2}^{i-1}(I - K_{i-m}C_{i-m})A_{i-m}(I - K_{0}C_{0})]
\end{align*}

\end{prop}
The proof is provided in Appendix A. 
\begin{remark}
    Proposition \ref{thm:1} formalizes the sensor measurements as $\y$ based on the augmented state variable $\tilde{\x}$ representing the system inputs. The importance of this new system model is it helps distinguish between the external inputs, which are deterministic assumed to be immune to false data injection, and the product state, which are impacted by the false data injection. We use this representation to estimate the augmented state variables and obtain the state estimation residuals, which will help detect and localize the attack.
\end{remark}
\begin{prop}\label{thm:2}
For an MMS described in section \ref{sec:3}, under a false data injection attack characterized by vector  ${\delta}=[{\delta_{0}}^T, \delta_1^T ... {\delta_K^T}]^T$, where $\delta_k$ represents the false data injected at stage $k$, the sensor measurement $\y$ can be expressed as
\begin{gather}
     \y = H \x + H_{1}{\delta},\\H_{1}= [\tilde{h}_{ij}]_{(K+1)\times(K+1)}\label{eq:y_attack}
 \end{gather}
where
\begin{gather}
  \tilde{h}_{ij}=
  \left\{
\begin{array}{lcl}
 {0}, & {i=1,..,j-1}\\
I,& {i=j}\\
C_{i-1}B_{i-1}L_{i-1}K_{j-1},&{i=j+1}\\
\end{array}\right\}
\end{gather}\\
for $i \geq j+2$:
\begin{align*}
\tilde{h}_{ij}= &C_{i-1}[\prod_{m=1}^{i-(j+1)}(A_{i-m} + B_{i-m}L_{i-m})B_{j}L_{j} +...\\
&\sum_{c=1}^{i-(j+2)}[\prod_{m=1}^{c}(A_{i-m} + B_{i-m}L_{i-m})B_{i-(c+1)}L_{i-(c+1)}... \\
&\prod_{b=c+2}^{i-j}(I - K_{i-b}C_{i-b})A_{i-b}] +...\\& B_{i-1}L_{i-1}\prod_{c=2}^{i-j}(I - K_{i-c}C_{i-c})A_{i-c}]K_{j}
\end{align*}
\end{prop}
The proof is provided in Appendix B.
\begin{remark}
In Proposition \ref{thm:2}, $H_{1}$ characterizes the relationship between the sensor measurements and the injected false data, with consideration of the attack propagation resulted from the multi-stage process. The equation \eqref{eq:y_attack} will facilitate extracting the important features from $\y$ to accurately detect FDI attacks and localize the source stage of the attack. Based on the linear relationship, we will develop the GLAHD framework based on the features extracted from the state estimation residuals.
\end{remark}

 To identify the anomalous pattern in the state estimation residuals, we need to analyse the variance of $\y$. In \eqref{eq:12}, as $\tilde{\x}$ only contains the input variable $\x_0$, while all the reference values are deterministic, the process noise and measurement noise $\w_k$ and $\v_k$ will contribute to the noise term $\epsilon$. Therefore, we derive the variance of $\epsilon$ in proposition 3, so that the patterns in $\epsilon$ can be used to identify the abnormal patterns introduced by the false data injected, $\delta$.
 
\begin{prop}\label{thm:3}
For an MMS described in section \ref{sec:3}, denote the covariance matrices of process and measurement noise as $\Sigma_x = diag(W_1,...,W_K)$ and $\Sigma_y = diag(V_1,...,V_K)$, respectively. The covariance of $\epsilon$ in \eqref{eq:12} follows:
\begin{gather}\label{eq:21}
\Sigma_\epsilon=H_{w}\Sigma_x H_{w}^T + H_1\Sigma_yH_1^T,
\end{gather}
where $H_{w}$ can be represented as: $H_{w}=[h_{ij}'']_{(K+1)\times(K+1)}$. In the above expression, for any $j$, we have:
    \begin{align*}
  h''_{ij}=
  \left\{
\begin{array}{lcl}
0, & {i=1,...,j-1}\\
 {C_{i-1}}, & {i=j}\\
{C_{i-1}[A_{i-1} + B_{i-1}L_{i-1}K_{i-2}C_{i-2}]},& {i=j+1}\\
\end{array}\right\}
   \end{align*}
 for $i \geq j+2$:
\begin{align*}
h''_{ij}= &C_{i-1}[\prod_{m=1}^{i-(j+1)}(A_{i-m} + B_{i-m}L_{i-m})(A_{j} +B_{j}L_{j}K_{j-1}C_{j-1}) ...\\
&\sum_{c=1}^{i-(j+2)}[\prod_{m=1}^{c}(A_{i-m} + B_{i-m}L_{i-m})B_{i-(c+1)}L_{i-(c+1)}... \\
&\prod_{b=c+2}^{i-j}(I - K_{i-b}C_{i-b})A_{i-b}(I - K_{j-1}C_{j-1})] + ... \\
&B_{i-1}L_{i-1}\prod_{c=2}^{i-j}(I - K_{i-c}C_{i-c})A_{i-c}(I - K_{j-1}C_{j-1})]
\end{align*}
\end{prop}
\begin{remark}
 Proposition \ref{thm:3} derives the expression of $\Sigma_\epsilon$, which will be used for feature extraction of the residuals $\epsilon$. By understanding the normal covariance matrix, the anomalous pattern in the data caused by the attack can be identified to detect the attack. The different anomalous patterns caused by attacks in different stages will be used to localize the attack.
\end{remark}

\subsection{The GLHAD Framework}\label{sub:4.2}
Based on the theoretical analysis, we propose a group Lasso-based hybrid attack detection (GLHAD) framework that incorporates the system dynamics into the group Lasso model. The GLAHD framework combines the advantage of both signature-based method and anomaly detection methods, where we can identify the location of the attack using a signature-based mechanism without relying on a comprehensive dataset tha contains labeled attack data to learn the features of attacks at different locations. Instead, it uses the theoretical analysis result to derive the signatures that associates with different attacks.

The GLAHD framework consists of two phases. In phase one, we estimate the system state and analyse the state estimation residuals to define the threshold of the detection algorithm. Under normal condition, the system model follows \eqref{eq:12}. The estimated augmented state variable, $\hat{\x}$, is calculated by projecting $\y$ onto the column space spanned by $H$:
\begin{gather*}
    \hat{\x}= (H^TH)^{-1}H^T\y,
\end{gather*}
and the state estimation residual, ${\r}$, is calculated as
 $${\r} = \y - \hat{\y},$$
 where
 $\r=
 \begin{bmatrix}
 {\r_{0}}^T & ... & {\r_K}^T
 \end{bmatrix}^T
 $ and ${\r_k}$ is the residuals of stage $k$.
  Intuitively, if $\r$ is close to 0, we can conclude that the system is not under attack. Hence, in phase two, we can apply the GLHAD framework to the residual to see whether or not the system is under attack. Also, in the case of an attack, the GLHAD framework provides us with enough information to detect the under-attack stage simultaneously. We must know that when we are projecting $\y$ into the spanned column space of $H$, some variation of $H_{1}$ is also explained by $H$. Hence, in the second phase, we replace $H_{1}$ with $R$. $R$ is the variation of $H_{1}$ which can not be explained by $H$. It is the difference between projection of $H_{1}$ onto the column space of $H$ and $H_{1}$:
$
R = H_{1} - H(H^TH)^{-1}H^TH_{1}.
$
We must know that the $R$ matrix may not be of full rank. Hence, we apply principal component analysis(PCA) to find vectors explaining the variance of $R$. The result can be seen in: $
R^{'} = PCA(R)
$. The system formulation in phase two is as follows:
  \begin{gather}\label{eq:24}
     {\r} = R^{'}{\delta}
 \end{gather}
In \eqref{eq:24}, we know the real values of {${\r}$} and $R^{'}$. One way to estimate the ${\delta}$ is by projecting ${\r}$ onto the column space spanned by $R^{'}$. Given only one stage can be under attack, we can conclude that in 
 \eqref{eq:y_attack} only one of the ${\delta_{i}}$ can be nonzero, ${\delta_{i}} \neq 0$. Hence, we can conclude that in \eqref{eq:y_attack}, different values of ${\delta}$ can be grouped based on the stage they belong to. In other words, referring to \eqref{eq:y_attack}, each ${\delta_{i}}$ can be considered a group. Regarding the \eqref{eq:24}, by considering the group structure of ${\delta}$, one of the best ways to estimate ${\delta}$ is by applying the group lasso regression algorithm. Originally, the group lasso algorithm was defined as regularised linear regression with the following loss function:
 \begin{gather}
 \min_{{\hat \delta}_{g} \in \mathbb{R^{'}}^{d_{g}}} \lVert{\sum_{g=0} ^K[R_{g}^{'}\hat{{\delta}}_{g}] - {\r}} \rVert_{2}^{2} + \lambda_{1}\sum_{g=0} ^K \sqrt{d_{g}}\lVert \hat{ {\delta}}_{g} \rVert_{2}^{2}\label{eq:25}\\
 \text{s.t.} \quad \y= H\hat{\x} + {\r} 
 \end{gather} 
 $R_{g}^{'} \in \R ^{N\times d_{g}}$ is the residual matrix corresponding to the covariance in stage $g$; in other words, it only includes the columns corresponding to stage $g$. ${\hat{\delta}_{g}}$ is the regression coefficients corresponding to group (stage) $g$. ${\r} \in \R^{N}$ is the residuals in \eqref{eq:24}. $N$ is the total number of sensor measurements, $N= \Sigma_{g=0} ^K n_{g} $, which $n_{g}$ is the number of sensors at stage $g$. $d_{g}$ is the dimensionality of stage (group) g, and $\lambda_{1}$ is the group-wise regularisation. In represented system, each group represents each stage of the system. We apply group lasso regression (GLR) on \eqref{eq:24} to estimate ${\delta}$. After applying GLR, the estimated coefficients are used to estimate ${ \r}$,
 $
     {\hat \r} = R{\hat \delta}
 $. Finally, for each stage, we apply the Hotteling $T^{2}$ test based on ${\hat \r}$ to see whether that stage is under attack or not. For applying Hotteling $T^{2}$ test, covariance matrix of ${\r}$ must be known. We know :
 $
 {\r} = (I - (H^TH)^{-1}H^T)\y.
 $
Hence, the covariance of ${\r}$ is:
 \begin{equation*}\label{eq:cov}
 Cov({\r}) = (I-(H^TH)^{-1}H^T)Cov(\y)(I-(H^TH)^{-1}H^T)^T
 \end{equation*}
 Same as matrix $R$, we must know that $Cov({\r})$ may not be of full rank. Hence, we apply eigendecomposition on the $Cov({\r})$ to extract meaningful eigenvalues and eigenvectors.
 \begin{gather}\label{eig_cov}
 Eig(vectors),Eig(values) = eigen- decomposition(Cov({\r}))
 \end{gather}
 The eigenvalues determine the covariance matrix. The new covariance matrix is a diagonal matrix comprising the calculated eigenvalues. It is calculated as $
Cov({\r})^{'}=diag(Eig(values))
 $. Also, since $Cov({\r})$ has changed, we must transfer the $\hat{{\r}}$ into the column space generated by the eigenvectors calculated in \eqref{eig_cov}. It is calculated as $
 \hat{{\r}}^{'} = \hat{{\r}}Eig(vectors)
 $. The main assumption for each stage is that the mean is zero, $\mu =0$. $T^{2}$ test statistic for stage $k$ is calculated as: 
 \begin{gather}
t^{2}_k= \hat{{{\r}}^{'}}_k^T{Cov({\r)_k}^{'}}^{-1}\hat{{\r}}_k^{'}
\end{gather}
${Cov({\r)_k}^{'}}^{-1}$ is the inverse covariance matrix of stage $k$.
For stage $k$, the control limit for $T^{2}$ test statistics is: $[0, UCL]$, where $UCL =\chi_{1-\alpha,d_k}^{2}$. $\alpha$ is type I error and $d_k$ is the dimension of stage $k$. Multiple stages may be under attack if more than one stage has test statistics outside the control limit. However, since only one stage can be under attack at a time, the stage with the maximum test statistics is considered the under-attack stage in such a scenario.
 The pseudo-code of the algorithm is given in Algorithm~\ref{al:1}:

\begin{algorithm}[h]
\caption{GL-based attack detection and identification for multistage linear system}\label{al:1}
\textbf{Input} \text{ :$H,H_{1}$, $\lambda_{1}$,$\alpha$, $\y$ }
\begin{algorithmic}[1]
    \State $\hat{\x} \gets (H^TH)^{-1}H^T\y$; $\hat{\y} \gets H\hat{\x}$;  ${\r} \gets \y-\hat{\y}$
    \State \text{Solve \eqref{eq:25}}
    
    \State $k \gets 0$, $pos \gets 0$, $max \gets 0$
    \While{$k \neq K$}
    \State $\hat{\r}_k \gets R{\hat{\delta}_k}$
    \State $t_k^2 \gets (\hat{\r}'_k)^T(\Sigma_\epsilon')_k^{-1}\hat{\r}'_k$   
    \If{$t^{2}_k > UCL$ \text{and} $t^{2}_k  \geq max$  }
    \State $max \gets {t}_k^{2}$
    \State $pos \gets k$

    \EndIf
    \State $k \gets k+1$
    \EndWhile

    \If{$pos > 0$}
    \State \Return{\text{stage of $pos$ is under attack}}

    \Else
    \State \Return{\text{no stage is under attack}}
    
    \EndIf

\end{algorithmic}
\end{algorithm}

\begin{figure}[ht]
    \centering
    \includegraphics[width=\linewidth]{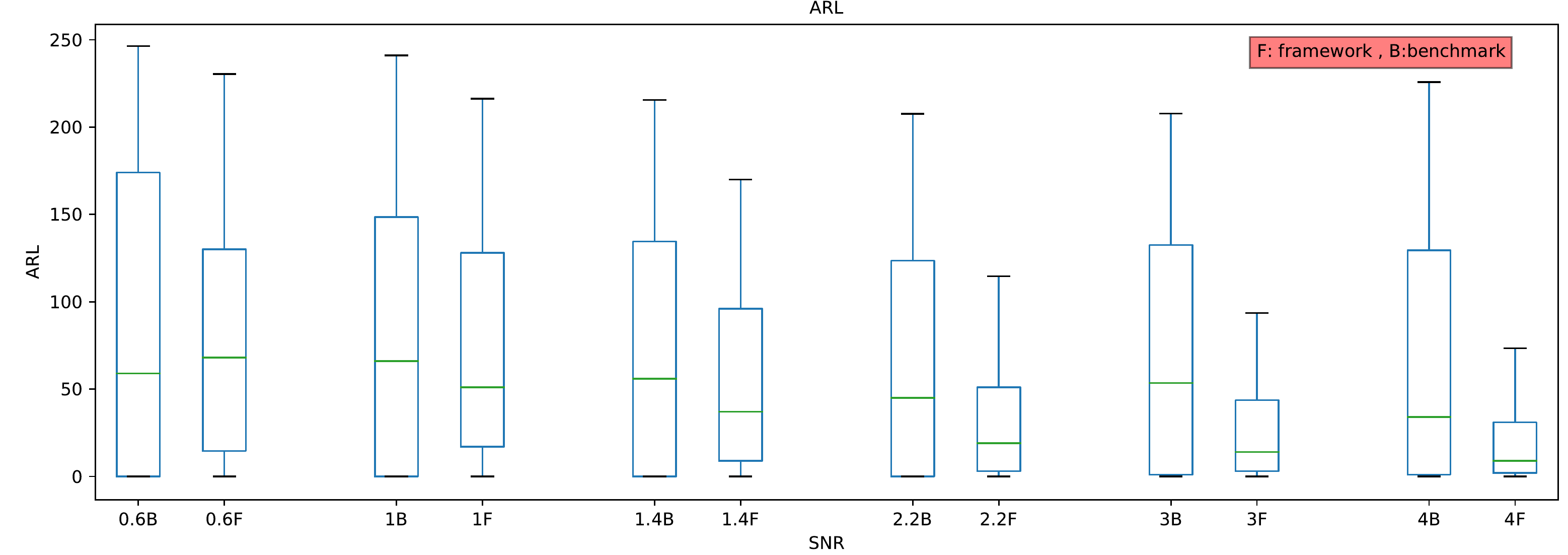}
\includegraphics[height=2in]{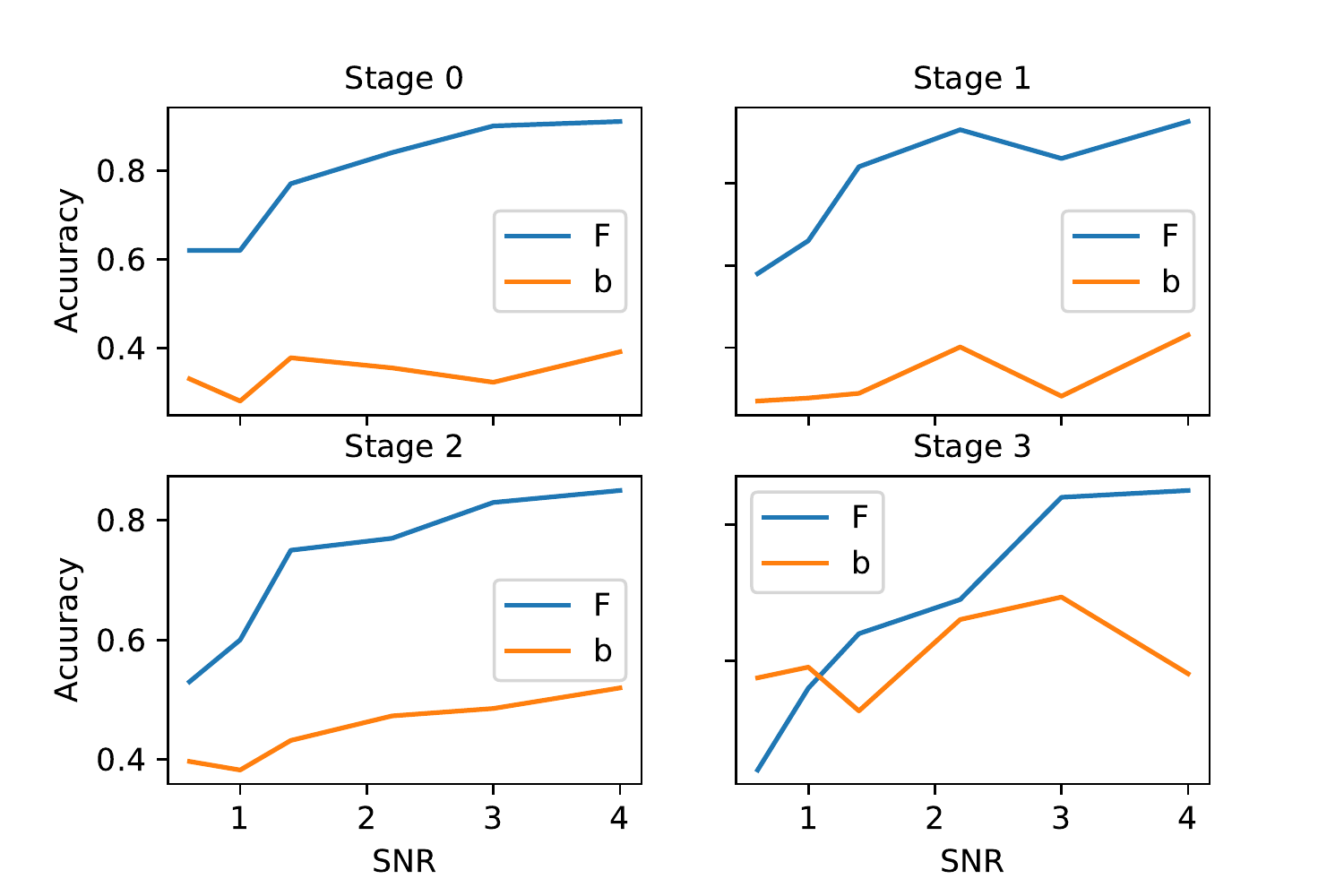}

        \caption{Numerical study results}

    \label{K-S-R}
\end{figure}

\section{Numerical Study}
This section compares the GLHAD framework with the in-stage $T^2$ test.  We randomly generate the elements in matrices ($ {A_{i}}_{3 \times 3}, {B_{i}}_{3 \times 3}, {C_{i}}_{5 \times 3}, i=0,1,2,3$) from a standard Gaussian distribution. Process noise and measurement noise at each stage comes from the multivariate normal distribution with mean zero and covariances: $ 
W = 0.1I_{3} , V =0.1I_{5}$. In the simulation study, we first run three hundred replications for the case when the system is not under attack. Then, we apply the GLHAD framework and benchmark method for normal data. Then, we simulate different sensor attacks on the system based on the different signal-to-noise ratios (SNR) and compare the performance of the two methods for attack detection. SNR is the magnitude of the simulated attack. It is defined as:
\begin{equation*}
SNR = \sqrt{{\delta_k^T}Cov(\y)_{kk}^{-1}{\delta_k}}
\end{equation*}
$Cov(\y)_{kk}^{-1}$ is the inverse covariance matrix of sensors at stage $k$. 
 We consider:  $  \x_0 = [1,1,1]^T $, $\x_{1}^r = {[-1.147,-0.726,-0.466]}^T$, $\x_{2}^r = {[0.239,-0.702,0.873]}^T $, $\x_{3}^r = {[0.108,-0.124,-0.140]}^T$.

\subsection{Stage-level $T^2$ test}\label{benchmark}
To evaluate the performance of GLAHD, we apply the in-stage $T^2$ test
as a benchmark method. Specifically, the residuals are obtained based on the measurement function:
\begin{gather}\label{eq:30}
\y = C\x
\end{gather}
In \eqref{eq:30}: $\label{eq:31}
C = diag(C_{0},...,C_K)
$. The estimated state variable, $\hat{\x}$, is calculated by projecting $\y$ into the column space spanned by $C$:
$
     \hat{\x} = (C^TC)^{-1}C^T\y.
 $ After calculation of $\hat{\x}$, the predicted measurements of the sensors, $\hat{\y}$, is calculated as:
 $
     \hat{\y} = H\hat{\x}
 $. Then, we calculate the residual of the algorithm: ${\epsilon_b} = \y - \hat{\y}$. The covariance matrix of ${\epsilon_b}$ can be calculated as:
\begin{equation*}\label{eq:cov_b}
 Cov({\epsilon_b}) = (I-(C^TC)^{-1}C^T)Cov(\y)(I-(C^TC)^{-1}C^T)^T
 \end{equation*}
Same as $Cov(\r)$, we must know that $Cov({\epsilon_b})$ may not be of full rank. Therefore, we apply eigendecomposition on the $Cov({\epsilon_b})$ to calculate eigenvalues and eigenvectors.
 \begin{gather}\label{eig_cov_b}
 Eig(vectors),Eig(values) = eigen- decomposition(Cov({\epsilon_b}))
 \end{gather}
 Again, Same as the GLHAD framework, The new covariance matrix is a diagonal matrix computed based on the calculated eigenvalues. It is calculated as $
Cov({\epsilon_b})^{'}=diag(Eig(values))
 $. Also, since $Cov({\epsilon_b})$ has changed, we must transfer the ${\epsilon_b}$ into the column space generated by the eigenvectors calculated in \eqref{eig_cov_b}. It is calculated as $
 {\r}^{'} = {\r}Eig(vectors)
 $. Then, we apply the $T^{2}$ test on each stage, similar to subsection \ref{sub:4.2}, to detect whether or not any stage is under attack. $T^{2}$ test statistic for stage $k$ is:
$
t^{2}_k= {\epsilon_{b_k}}^TCov({{\r}^{'}_b)_k^{-1}}{\epsilon_{b_k}}
$. ${Cov({\epsilon_b)_k}^{'}}^{-1}$ is the inverse covariance matrix of $Cov({\epsilon_b)^{'}}$ at stage $k$. For stage $k$, the control limit for $T^{2}$ test statistics is: $[0, UCL]$, where $UCL =\chi_{1-\alpha,d_k}^{2}$. $\alpha$ is type I error and $d_k$ is the dimension of stage $k$. Again, similar to the GLHAD framework, multiple stages may be under-attack if more than one stage has test statistics outside the control limit. However, since only one stage can be under attack at a time, the stage with the maximum test statistics is considered the under-attack stage in such a scenario.


\begin{figure}[t]
    \centering
\includegraphics[width=\linewidth]{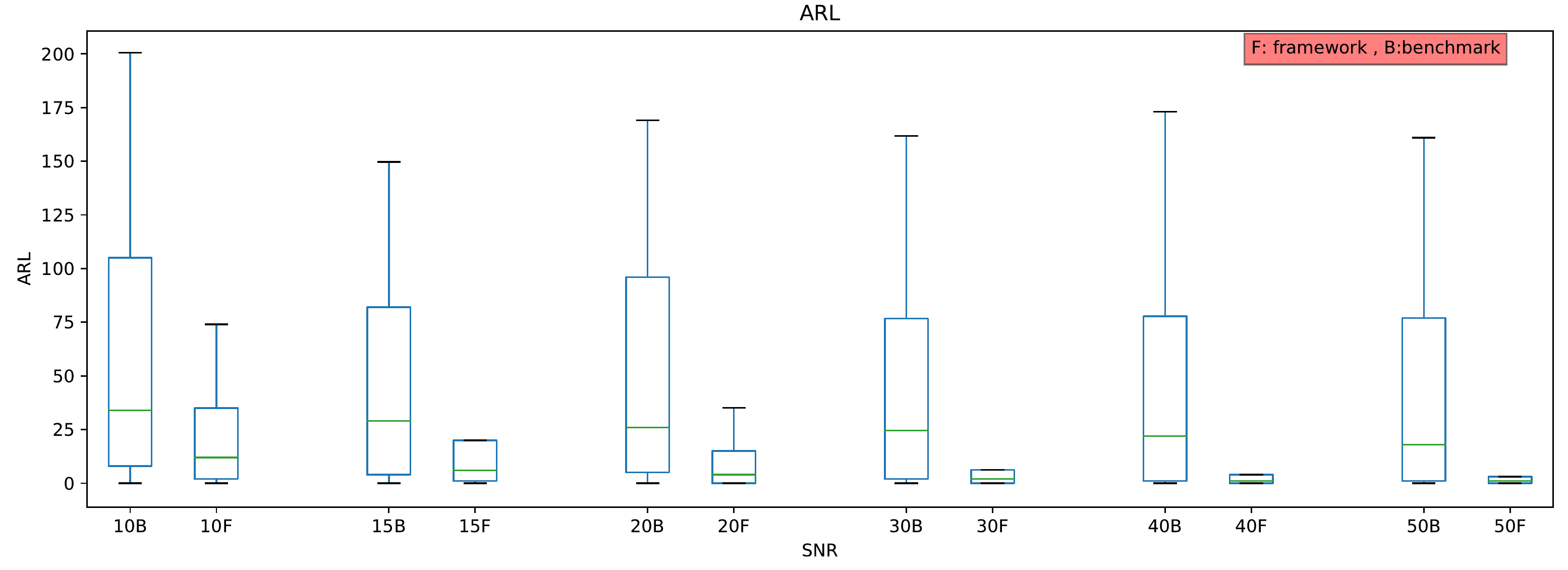}
\includegraphics[height=2in]{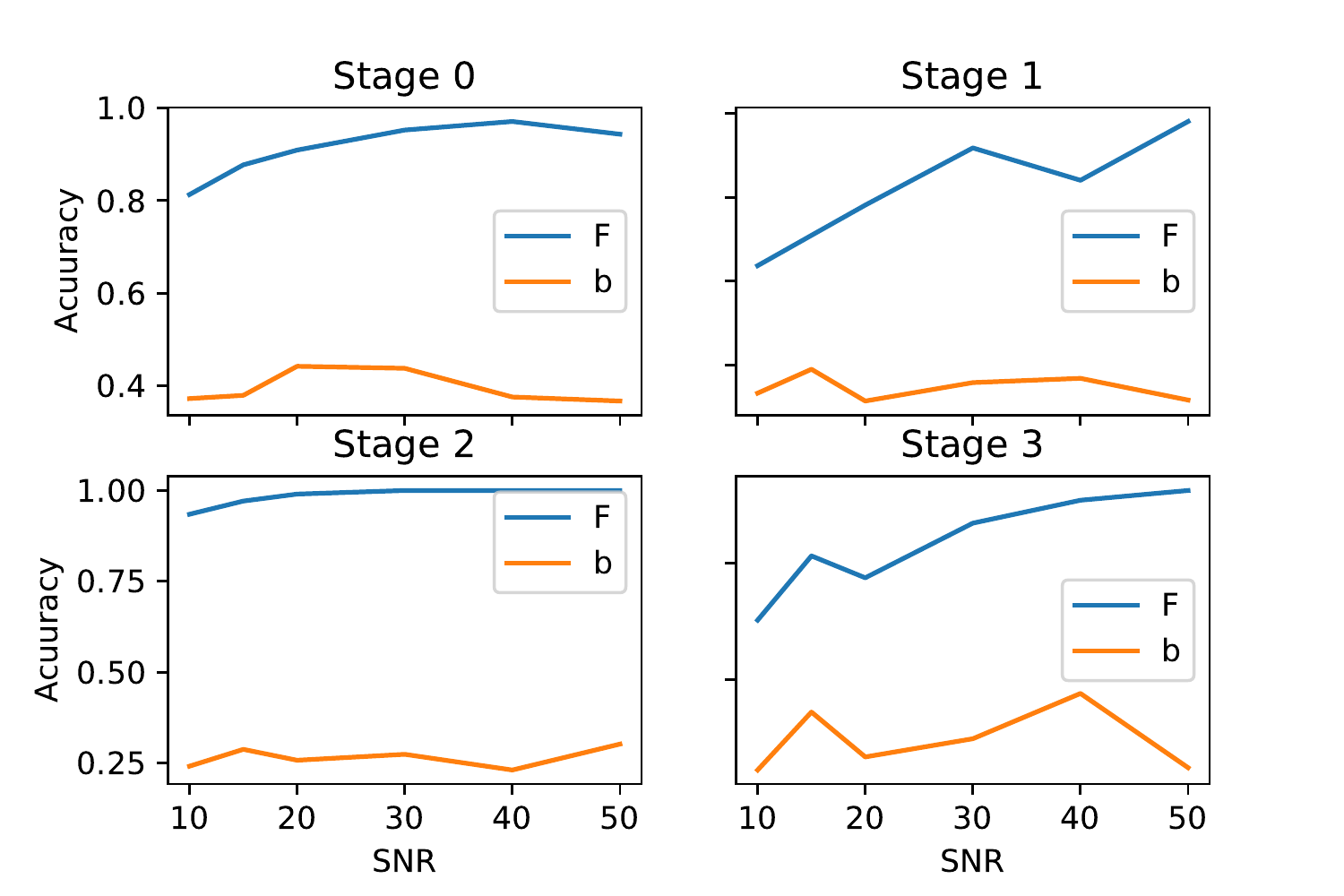}
        \caption{Case study results}
    \label{C-S-R}
\end{figure}


\begin{table}[t]
    \centering
\includegraphics[width=\linewidth]{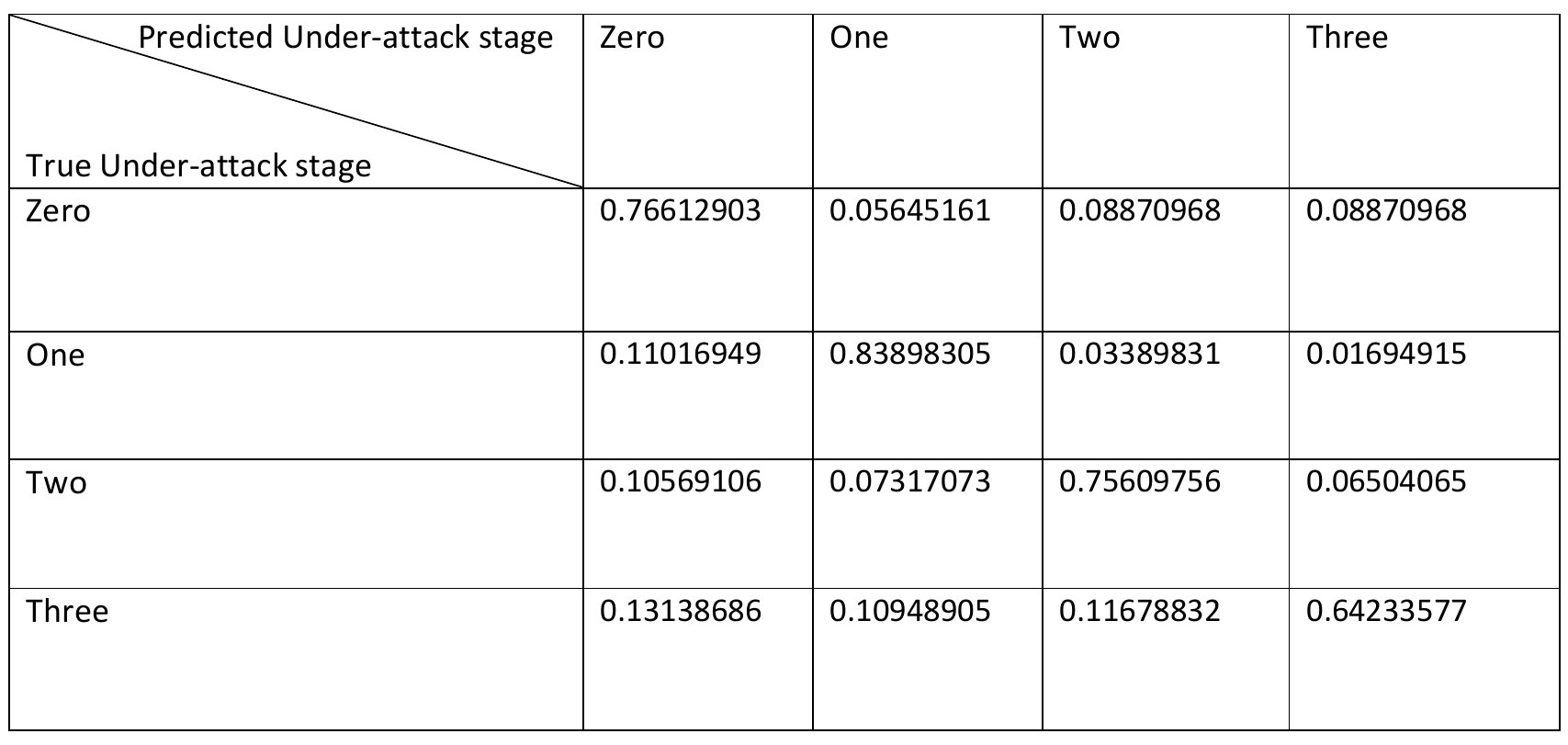}
        \caption{GLHAD framework}
    \label{Table_GLHAD}
\end{table}
\begin{table}[t]
    \centering
\includegraphics[width=\linewidth]{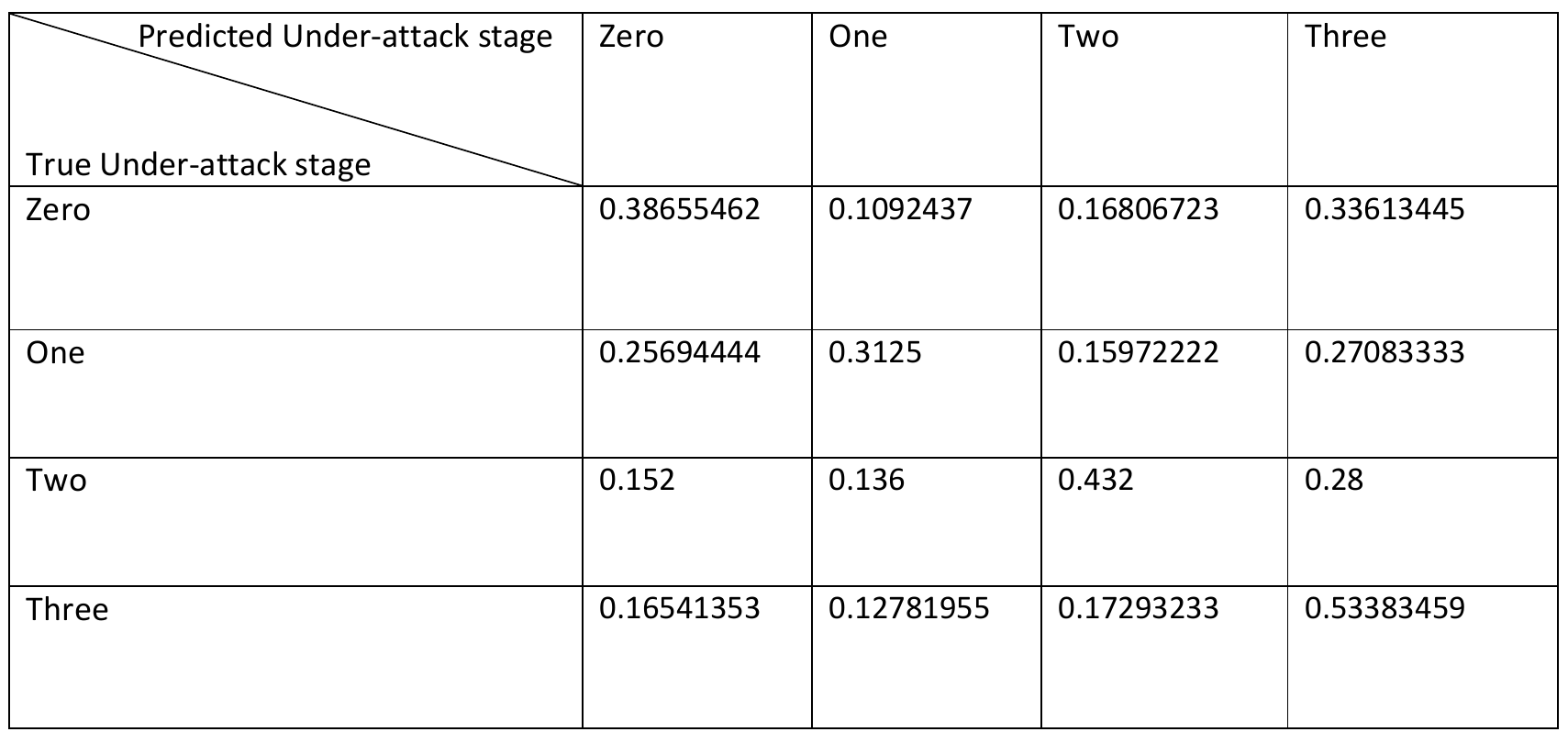}
        \caption{Benchmark method}
    \label{Table_bench}
\end{table}


\begin{table}[t]
    \centering
\includegraphics[width=\linewidth]{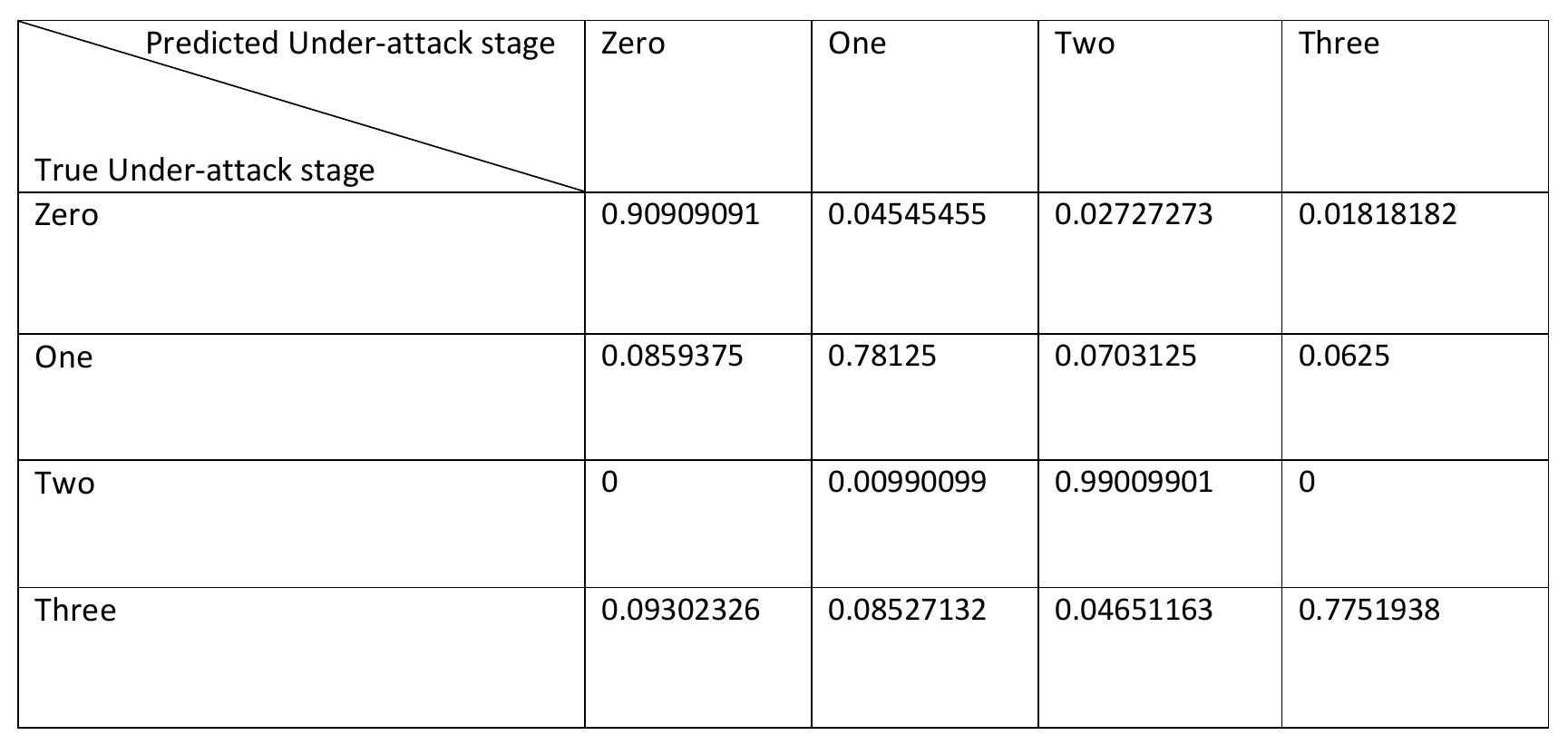}
        \caption{GLHAD framework}
    \label{Table_GLHAD_real}
\end{table}
\begin{table}[t]
    \centering
\includegraphics[width=\linewidth]{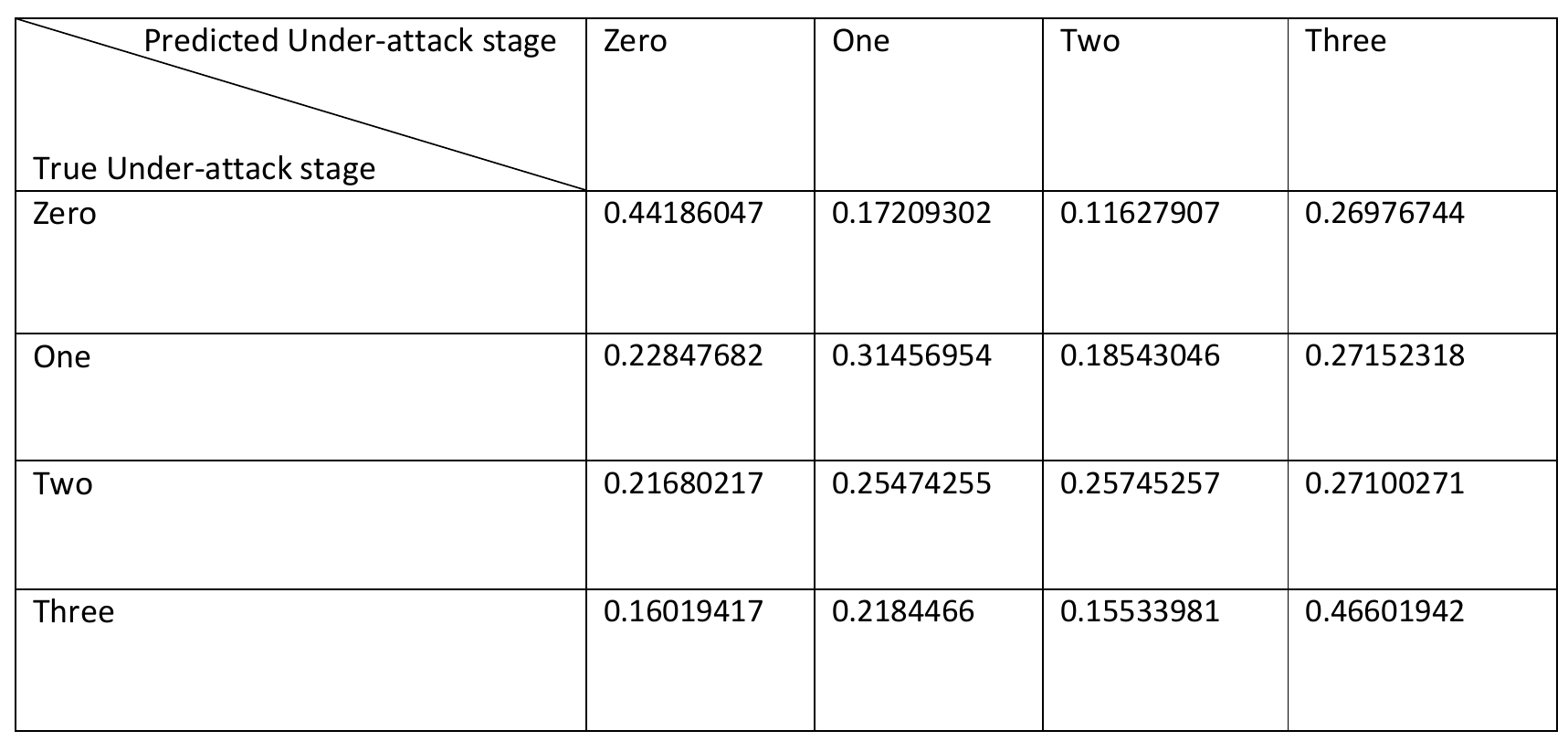}
        \caption{Benchmark method}
    \label{Table_bench_real}
\end{table}

\subsection{Numerical results}

 For each algorithm, two criteria are calculated: $(i)$ average run length (ARL): ARL represents the average number of samples a method needs before an out-of-control condition is detected. 
 $(ii)$ true localization: percentage of samples correctly detecting the under-attack stage. For the ARL and true localization, the results for two algorithms are calculated for SNR 
 $\in \{0.6,1,1.4,2.2,3,4\}$. The results are shown in Fig. \ref{K-S-R}. In this figure, "F" stands for the GLHAD framework and "b" stands for the benchmark method. Regarding the ARL, the GLHAD framework needs  fewer samples to detect attacks than the benchmark method, except for SNR = $0.6$. Also, for other quartiles, the GLHAD outperforms the benchmark method. For true localization, we compare both methods stage by stage. For example, when stage $1$ is under attack, we calculate the accuracy of correctly attributing the attack to stage $1$ for both the GLHAD framework and benchmark method. It can be seen that for stage $k$, $k \in \{0,1,2\}$, for all the values of SNR, the GLHAD framework correctly localizes the attack with higher accuracy compared to the benchmark method. For stage $3$, the benchmark method has higher accuracy for SNR $=0.6$, but for all the other values of SNR, the GLHAD framework has higher accuracy than the benchmark method. For true localization, we also describe the localization detail for SNR$=1.4$. The details can be seen in Fig.~\ref{Table_GLHAD} and Fig.~\ref{Table_bench}. In these two figures, the rows represent the truly under-attack stage, and the columns represent the predicted under-attack stage. Finally, based on the results of the two criteria, we can conclude that the GLHAD framework outperforms the benchmark method.

\subsection{Case study }
In this study, we implement the GLHAD framework and benchmark method on a real-world multistage assembly process case study to evaluate the performance of our algorithm. To extract real-world data, we use defining matrices $(A_{i},B_{i}, i=1,2,3)$ from \cite{shi2006stream} and select only the first six rows and columns of the matrices. It is essential to note that large matrix values can lead to increased covariance, resulting in an unstable system. To stabilize the system, we multiply the matrices by $0.01$.

Next, we apply the GLHAD framework and benchmark method to the extracted data. The ARLs of the baseline method group-wise hypothesis test and the proposed GLHAD framework are shown in Fig.~\ref{C-S-R}. In this figure, "F" represents the GLHAD framework, while "b" denotes the benchmark method. We test the attacks for SNR values in the set ${10,15,20,30,40,50}$. The results show that GLHAD framework generates a lower ARL, which means it outperforms the benchmark method for all SNR values. It also demonstrates a better performance in terms of the variance of the ARL. In terms of true localization, the GLHAD framework exhibits higher accuracy than the benchmark method for stages zero, one, and two across all SNR values, except for stage zero, where the benchmark method outperforms GLHAD for SNR values of $10,15,20$.

Additionally, we provide detailed localization results for an SNR value of $20$, which can be found in Fig.\ref{Table_GLHAD_real} and Fig.\ref{Table_bench_real}. In these figures, rows represent the truly under-attack stage, and columns represent the predicted under-attack stage.

The results show that the GLHAD framework demonstrates superior performance compared to the benchmark method, indicating its effectiveness for detecting and localizing cyberattacks in multistage assembly processes using real-world data.

\section{Conclusion}

This paper proposes a new system representation for the MMSs. We consider a general model for MMS, which is unprecedented for attack detection. Also, we propose the GLHAD framework for detecting cyber attacks on the sensors. This algorithm can detect the attack and localize it to the correct stage simultaneously with high accuracy. This framework can be easily generalized for any MMS. For future research, we aim to work on designing new system representations, including other types of attacks in the MMS, and try developing new algorithms for detecting them.

\bibliographystyle{asmems4}

\bibliography{asme2e}

\begin{thebibliography}{10}

\bibitem{mahoney2017cybersecurity}
Mahoney, T.~C., and Davis, J., 2017.
\newblock Cybersecurity for manufacturers: Securing the digitized and connected
  factory.
\newblock Tech. rep.

\bibitem{langner2011stuxnet}
Langner, R., 2011.
\newblock ``Stuxnet: Dissecting a cyberwarfare weapon''.
\newblock {\em IEEE Security \& Privacy, {\bf 9}}(3), pp.~49--51.

\bibitem{wu2019intrusion}
Wu, M., 2019.
\newblock ``Intrusion detection for cyber-physical attacks in
  cyber-manufacturing system''.
\newblock PhD thesis, Syracuse University.

\bibitem{lee2014german}
Lee, R.~M., Assante, M.~J., and Conway, T., 2014.
\newblock ``German steel mill cyber attack''.
\newblock {\em Industrial Control Systems, {\bf 30}}(62), pp.~1--15.

\bibitem{singh2020review}
Singh, S., Yadav, N., and Chuarasia, P.~K., 2020.
\newblock ``A review on cyber physical system attacks: Issues and challenges''.
\newblock In 2020 International Conference on Communication and Signal
  Processing (ICCSP), IEEE, pp.~1133--1138.

\bibitem{liu2021cyber}
Liu, T., Yang, B., Li, Q., Ye, J., Song, W., and Liu, P., 2021.
\newblock ``Cyber-physical taint analysis in multi-stage manufacturing systems
  (mms): A case study''.
\newblock {\em arXiv preprint arXiv:2109.12774}.

\bibitem{shi2006stream}
Shi, J., 2006.
\newblock {\em Stream of variation modeling and analysis for multistage
  manufacturing processes}.
\newblock CRC press.

\bibitem{al2022securing}
Al~Mamun, A., Liu, C., Kan, C., and Tian, W., 2022.
\newblock ``Securing cyber-physical additive manufacturing systems by in-situ
  process authentication using streamline video analysis''.
\newblock {\em Journal of Manufacturing Systems, {\bf 62}}, pp.~429--440.

\bibitem{shi2022lstm}
Shi, Z., Mamun, A.~A., Kan, C., Tian, W., and Liu, C., 2022.
\newblock ``An lstm-autoencoder based online side channel monitoring approach
  for cyber-physical attack detection in additive manufacturing''.
\newblock {\em Journal of Intelligent Manufacturing}, pp.~1--17.

\bibitem{zeltmann2016manufacturing}
Zeltmann, S.~E., Gupta, N., Tsoutsos, N.~G., Maniatakos, M., Rajendran, J., and
  Karri, R., 2016.
\newblock ``Manufacturing and security challenges in 3d printing''.
\newblock {\em Jom, {\bf 68}}(7), pp.~1872--1881.

\bibitem{liu2020online}
Liu, C., Kan, C., and Tian, W., 2020.
\newblock ``An online side channel monitoring approach for cyber-physical
  attack detection of additive manufacturing''.
\newblock In International Manufacturing Science and Engineering Conference,
  Vol.~84263, American Society of Mechanical Engineers, p.~V002T07A016.

\bibitem{liu2011false}
Liu, Y., Ning, P., and Reiter, M.~K., 2011.
\newblock ``False data injection attacks against state estimation in electric
  power grids''.
\newblock {\em ACM Transactions on Information and System Security (TISSEC),
  {\bf 14}}(1), pp.~1--33.

\bibitem{northern2021vercasm}
Northern, B., Burks, T., Hatcher, M., Rogers, M., and Ulybyshev, D., 2021.
\newblock ``Vercasm-cps: Vulnerability analysis and cyber risk assessment for
  cyber-physical systems''.
\newblock {\em Information, {\bf 12}}(10), p.~408.

\bibitem{zhang2022modeling}
Zhang, Y., Jiang, T., Shi, Q., Liu, W., and Huang, S., 2022.
\newblock ``Modeling and vulnerability assessment of cyber physical system
  considering coupling characteristics''.
\newblock {\em International Journal of Electrical Power \& Energy Systems,
  {\bf 142}}, p.~108321.

\bibitem{pan2020modeling}
Pan, H., Lian, H., Na, C., and Li, X., 2020.
\newblock ``Modeling and vulnerability analysis of cyber-physical power systems
  based on community theory''.
\newblock {\em IEEE Systems Journal, {\bf 14}}(3), pp.~3938--3948.

\bibitem{pivoto2021cyber}
Pivoto, D.~G., de~Almeida, L.~F., da~Rosa~Righi, R., Rodrigues, J.~J., Lugli,
  A.~B., and Alberti, A.~M., 2021.
\newblock ``Cyber-physical systems architectures for industrial internet of
  things applications in industry 4.0: A literature review''.
\newblock {\em Journal of manufacturing systems, {\bf 58}}, pp.~176--192.

\bibitem{patan2020smart}
Patan, R., Ghantasala, G.~P., Sekaran, R., Gupta, D., and Ramachandran, M.,
  2020.
\newblock ``Smart healthcare and quality of service in iot using grey filter
  convolutional based cyber physical system''.
\newblock {\em Sustainable Cities and Society, {\bf 59}}, p.~102141.

\bibitem{thakur2021intrusion}
Thakur, S., Chakraborty, A., De, R., Kumar, N., and Sarkar, R., 2021.
\newblock ``Intrusion detection in cyber-physical systems using a generic and
  domain specific deep autoencoder model''.
\newblock {\em Computers \& Electrical Engineering, {\bf 91}}, p.~107044.

\bibitem{althobaiti2021intelligent}
Althobaiti, M.~M., Kumar, K. P.~M., Gupta, D., Kumar, S., and Mansour, R.~F.,
  2021.
\newblock ``An intelligent cognitive computing based intrusion detection for
  industrial cyber-physical systems''.
\newblock {\em Measurement, {\bf 186}}, p.~110145.

\bibitem{kazemi2021finite}
Kazemi, Z., Safavi, A.~A., Arefi, M.~M., and Naseri, F., 2021.
\newblock ``Finite-time secure dynamic state estimation for cyber--physical
  systems under unknown inputs and sensor attacks''.
\newblock {\em IEEE Transactions on Systems, Man, and Cybernetics: Systems,
  {\bf 52}}(8), pp.~4950--4959.

\bibitem{ding2020secure}
Ding, D., Han, Q.-L., Ge, X., and Wang, J., 2020.
\newblock ``Secure state estimation and control of cyber-physical systems: A
  survey''.
\newblock {\em IEEE Transactions on Systems, Man, and Cybernetics: Systems,
  {\bf 51}}(1), pp.~176--190.

\bibitem{zhao2022anti}
Zhao, Y., Du, X., Zhou, C., and Tian, Y.-C., 2022.
\newblock ``Anti-saturation resilient control of cyber-physical systems under
  actuator attacks''.
\newblock {\em Information Sciences, {\bf 608}}, pp.~1245--1260.

\bibitem{liao2013intrusion}
Liao, H.-J., Lin, C.-H.~R., Lin, Y.-C., and Tung, K.-Y., 2013.
\newblock ``Intrusion detection system: A comprehensive review''.
\newblock {\em Journal of Network and Computer Applications, {\bf 36}}(1),
  pp.~16--24.

\bibitem{yaacoub2020cyber}
Yaacoub, J.-P.~A., Salman, O., Noura, H.~N., Kaaniche, N., Chehab, A., and
  Malli, M., 2020.
\newblock ``Cyber-physical systems security: Limitations, issues and future
  trends''.
\newblock {\em Microprocessors and microsystems, {\bf 77}}, p.~103201.

\bibitem{panigrahi2022intrusion}
Panigrahi, R., Borah, S., Pramanik, M., Bhoi, A.~K., Barsocchi, P., Nayak,
  S.~R., and Alnumay, W., 2022.
\newblock ``Intrusion detection in cyber--physical environment using hybrid
  na{\"\i}ve bayes—decision table and multi-objective evolutionary feature
  selection''.
\newblock {\em Computer Communications, {\bf 188}}, pp.~133--144.

\bibitem{kwon2022advanced}
Kwon, H.-Y., Kim, T., and Lee, M.-K., 2022.
\newblock ``Advanced intrusion detection combining signature-based and
  behavior-based detection methods''.
\newblock {\em Electronics, {\bf 11}}(6), p.~867.

\bibitem{song2020layered}
Song, J., Bandaru, H., He, X., Qiu, Z., and Moon, Y.~B., 2020.
\newblock ``Layered image collection for real-time defective inspection in
  additive manufacturing''.
\newblock In ASME International Mechanical Engineering Congress and Exposition,
  Vol.~84492, American Society of Mechanical Engineers, p.~V02BT02A006.

\bibitem{wu2016detecting}
Wu, M., Phoha, V.~V., Moon, Y.~B., and Belman, A.~K., 2016.
\newblock ``Detecting malicious defects in 3d printing process using machine
  learning and image classification''.
\newblock In ASME International Mechanical Engineering Congress and Exposition,
  Vol.~50688, American Society of Mechanical Engineers, p.~V014T07A004.

\bibitem{song2019physical}
Song, J., Shukla, D., Wu, M., Phoha, V.~V., and Moon, Y.~B., 2019.
\newblock ``Physical data auditing for attack detection in cyber-manufacturing
  systems: Blockchain for machine learning process''.
\newblock In ASME International Mechanical Engineering Congress and Exposition,
  Vol.~59384, American Society of Mechanical Engineers, p.~V02BT02A004.

\bibitem{wu2019detecting}
Wu, M., Song, Z., and Moon, Y.~B., 2019.
\newblock ``Detecting cyber-physical attacks in cybermanufacturing systems with
  machine learning methods''.
\newblock {\em Journal of intelligent manufacturing, {\bf 30}}(3),
  pp.~1111--1123.

\bibitem{bhardwaj2020capturing}
Bhardwaj, A., Al-Turjman, F., Kumar, M., Stephan, T., and Mostarda, L., 2020.
\newblock ``Capturing-the-invisible (cti): Behavior-based attacks recognition
  in iot-oriented industrial control systems''.
\newblock {\em IEEE access, {\bf 8}}, pp.~104956--104966.

\bibitem{qian2020cyber}
Qian, J., Du, X., Chen, B., Qu, B., Zeng, K., and Liu, J., 2020.
\newblock ``Cyber-physical integrated intrusion detection scheme in scada
  system of process manufacturing industry''.
\newblock {\em IEEE Access, {\bf 8}}, pp.~147471--147481.

\bibitem{abokifa2019real}
Abokifa, A.~A., Haddad, K., Lo, C., and Biswas, P., 2019.
\newblock ``Real-time identification of cyber-physical attacks on water
  distribution systems via machine learning--based anomaly detection
  techniques''.
\newblock {\em Journal of Water Resources Planning and Management, {\bf
  145}}(1), p.~04018089.

\bibitem{urbina2016limiting}
Urbina, D.~I., Giraldo, J.~A., Cardenas, A.~A., Tippenhauer, N.~O., Valente,
  J., Faisal, M., Ruths, J., Candell, R., and Sandberg, H., 2016.
\newblock ``Limiting the impact of stealthy attacks on industrial control
  systems''.
\newblock In Proceedings of the 2016 ACM SIGSAC conference on computer and
  communications security, pp.~1092--1105.

\bibitem{li2020detection}
Li, D., Gebraeel, N., and Paynabar, K., 2020.
\newblock ``Detection and differentiation of replay attack and equipment faults
  in scada systems''.
\newblock {\em IEEE Transactions on Automation Science and Engineering, {\bf
  18}}(4), pp.~1626--1639.

\bibitem{mo2013detecting}
Mo, Y., Chabukswar, R., and Sinopoli, B., 2013.
\newblock ``Detecting integrity attacks on scada systems''.
\newblock {\em IEEE Transactions on Control Systems Technology, {\bf 22}}(4),
  pp.~1396--1407.

\bibitem{van2015sequential}
Van~Long, D., FILLATRE, L., and NIKIFOROV, I., 2015.
\newblock ``Sequential monitoring of scada systems against cyber/physical
  attacks''.
\newblock {\em IFAC-PapersOnLine, {\bf 48}}(21), pp.~746--753.

\bibitem{mo2009secure}
Mo, Y., and Sinopoli, B., 2009.
\newblock ``Secure control against replay attacks''.
\newblock In 2009 47th annual Allerton conference on communication, control,
  and computing (Allerton), IEEE, pp.~911--918.

\bibitem{li2019feasibility}
Li, B., Xiao, G., Lu, R., Deng, R., and Bao, H., 2019.
\newblock ``On feasibility and limitations of detecting false data injection
  attacks on power grid state estimation using d-facts devices''.
\newblock {\em IEEE Transactions on Industrial Informatics, {\bf 16}}(2),
  pp.~854--864.

\bibitem{wang2019review}
Wang, Q., Tai, W., Tang, Y., and Ni, M., 2019.
\newblock ``Review of the false data injection attack against the
  cyber-physical power system''.
\newblock {\em IET Cyber-Physical Systems: Theory \& Applications, {\bf 4}}(2),
  pp.~101--107.

\bibitem{jorjani2020graph}
Jorjani, M., Seifi, H., and Varjani, A.~Y., 2020.
\newblock ``A graph theory-based approach to detect false data injection
  attacks in power system ac state estimation''.
\newblock {\em IEEE Transactions on Industrial Informatics, {\bf 17}}(4),
  pp.~2465--2475.

\end{thebibliography}

\appendix       
\section*{Appendix A: Proof of Proposition \ref{thm:1}}\label{Ap:A}
We first consider the scenario when $j \in \{2,3,...,K\}$. We present a proof for $j=2$. This proof can be easily generalized for any other $j \in \{3,..., K\}$. $h_{i2}$ represents the coefficient of $\x_{1}^r$ in $ \y_{i-1}$.
It can be easily seen that : $h_{12}=0$ , $h_{22} = C_{1}B_{1}L_{R_{1}}$.
 For $n\geq2$, we use induction to complete the proof. For $i=3$, $h_{32}$ represents the coefficient of $\x_{1}^T$ in $\y_{2}$ and we have:
 ${h}_{32} ={C}_{2}[({A}_{2}+{B}_{2}{L}_{2}]B_{1}L_{R_{1}}
 $. We assume that for an arbitrary $i\geq 3$:
 \begin{equation*}{h}_{i2}={C}_{i-1}\prod_{b=2}^{i-1}({A}_{b}+{B}_{b}{L}_{b})B_{1}L_{R_{1}}
 \end{equation*} We must show that:  $
{h}_{(i+1)2}={C}_{i}\prod_{b=2}^{i}({A}_{b}+{B}_{b}{L}_{b})B_{1}L_{R_{1}}
$. From the assumption, it can be seen that:
$
\gamma_{1,i-2}=\prod_{b=2}^{i-2}({A}_{b}+{B}_{b}{L}_{b})B_{1}L_{R_{1}}
$, where $\gamma_{i,j}$ is the  The coefficient of $x_{i}^r$ in $hat{\x}_{j|j}$. Hence, we can conclude that:
\begin{align*}
\gamma_{1,i-1}&= (I - K_{i-1}C_{i-1})(A_{i-1} + B_{i-1}L_{i-1})j_{2} + K_{i-1}h_{i2}\\
&= (A_{i-1} + B_{i-1}L_{i-1})j_{2}\\
&=\prod_{b=2}^{i-1}({A}_{b}+{B}_{b}{L}_{b})B_{1}L_{R_{1}}
\end{align*}
We know that :
$
\y_{i} = {C}_{i}\x_{i}= {C}_{i}{A}_{i}\x_{i-1} + {C}_{i}{B}_{i}\u_{i}= {C}_{i}{A}_{i}\x_{i-1} + {C}_{i}{B}_{i}[L_{i}\hat{\x}_{i-1|i-1}+L_{R_{i}}x_{i}^T] 
$. From assumption, it can be seen that:
$\beta_{1,i-1}= \prod_{b=2}^{i-1}({A}_{b}+{B}_{b}{L}_{b})B_{1}L_{R_{1}}
 $. $\beta_{i,j}$ represents the coefficient of $x_{i}^r$ in $x_{j}$. Hence, we can conclude that: 
\begin{equation*}
 h_{(i+1)2}={C}_{i}{A}_{i}\beta_{1,i-1} + {C}_{i}{B}_{i}{L}_{i}\gamma_{1,i-1}
 = {C}_{i}[\prod_{b=2}^{i}({A}_{b}+{B}_{b}{L}_{b})B_{1}L_{R_{1}}]
\end{equation*}
 
When $j=1$, $h_{i1}$ represents the coefficient of $\x_0$ in $\y_{i-1}$. Hence, $h_{11}$ and $h_{12}$ represents the coefficient of $\x_0$ in ${y_{0}}$ and ${y_{1}}$, respectively. We use induction for the proof for $i\geq3$.\newline
$h_{31}$ represents the coefficient of $\x_0$ in $\y_{2}$, so we have :
$
h_{31} = {C}_{2}[({A}_{2}+{B}_{2}{L}_{2})(A_{1} + B_{1}L_{1}K_{0}C_{0}) - B_{2}L_{2}(I - K_{1}C_{1})A_{1}(I - K_{0}C_{0})]
$. We assume that for an arbitrary $i \geq 3$, we have :
\begin{align*}
h_{i1} = &C_{i-1}[\prod_{k=1}^{i-2}(A_{i-k} + B_{i-k}L_{i-k})(A_{1} + B_{1}L_{1}K_{0}C_{0}) - ...
\\&              \sum_{j=1}^{i-3} [\prod_{k=1}^{j}(A_{i-k} + B_{i-k}L_{i-k})(B_{i-(j+1)}L_{i-(j+1)})...\\
&\prod_{c=j+2}^{i-1}(I - K_{i-c}C_{i-c})A_{i-c}(I - K_{0}C_{0})] - (B_{i-1}L_{i-1})...\\
&\prod_{m=2}^{i-1}(I - K_{i-m}C_{i-m})A_{i-m}(I - K_{0}C_{0})]
\end{align*}

From the assumption, it can be seen that: 
\begin{align*}
\theta_{i-2,i-1} = &\prod_{k=2}^{i-2}(A_{i-k} + B_{i-k}L_{i-k})(A_{1} + B_{1}L_{1}K_{0}C_{0}) - ...\\&         \sum_{j=1}^{i-3} [\prod_{k=2}^{j}(A_{i-k} + B_{i-k}L_{i-k})(B_{i-(j+1)}L_{i-(j+1)})...\\
&\prod_{c=j+2}^{i-1}(I - K_{i-c}C_{i-c})A_{i-c}(I - K_{0}C_{0})]  -...\\
&\prod_{m=2}^{i-1}(I - K_{i-m}C_{i-m})A_{i-m}(I - K_{0}C_{0})
\end{align*}
Where $\theta_{i,j}$ is the coefficient of $\x_0$ of $\hat{\x}_{i|i}$ in ${y_{j}}$. Also, it can be seen that :
\begin{align*}
\pi_{i-1} = &\prod_{k=1}^{i-2}(A_{i-k} + B_{i-k}L_{i-k})(A_{1} + B_{1}L_{1}K_{0}C_{0}) - ...\\             
&sum_{j=1}^{i-3} [\prod_{k=1}^{j}(A_{i-k} + B_{i-k}L_{i-k})(B_{i-(j+1)}L_{i-(j+1)})... \\
&\prod_{c=j+2}^{i-1}(I - K_{i-c}C_{i-c})A_{i-c}(I - K_{0}C_{0})]  -...\\
&(B_{i-1}L_{i-1})\prod_{m=2}^{i-1}(I - K_{i-m}C_{i-m})A_{i-m}(I - K_{0}C_{0})
\end{align*}
$\pi_{i}$ is the coefficient of $\x_0$ in ${x_{i}}$.
The coefficient of $\x_0$ in ${y_{i}}$ is : 
 \begin{align*}
h_{(i+1)1}&=C_{i}[A_{i}\pi_{i-1} + B_{i}L_{i}((I - K_{i-1}C_{i-1})...\\&(A_{i-1} + B_{i-1}L_{i-1})\theta_{i-2,i-1} - K_{i-1}{y_{i-1}} )]
\end{align*}
It can be seen that:
\begin{align*}
&(I - K_{i-1}C_{i-1})(A_{i-1} + B_{i-1}L_{i-1})\theta_{i-2,i-1} - K_{i-1}{y_{i-1}}\\=&
(A_{i-1} + B_{i-1}L_{i-1})j_{1} + K_{i-1}C_{i-1}\prod_{b=1}^{i-1}A_{i-b}(I - K_{i-b-1}C_{i-b-1})
\end{align*}
Hence, the coefficient of $\x_0$ in ${y_{i}}$ is:

\begin{align*}
&h_{(i+1)1}\\=&C_{i}[A_{i}\pi_{i-1} + B_{i}L_{i}((I - K_{i-1}C_{i-1})(A_{i-1} + B_{i-1}L_{i-1})\theta_{i-2,i-1}- K_{i-1}{y_{i-1}} )]\\=&C_{i}[A_{i}\pi_{i-1} + B_{i}L_{i}((A_{i-1} + B_{i-1}L_{i-1})\theta_{i-2,i-1} + K_{i-1}C_{i-1}...\\&\prod_{b=1}^{i-1}A_{i-b}(I - K_{i-b-1}C_{i-b-1})
\end{align*}
The above statement is precisely what we must show. The proof is complete.

\section*{Proof of Propositions 2 and 3}
Propositions 2 and 3 can be proven by induction. Regarding Proposition $2$, for $i= 1,...,j+1$, the proof is obvious. For $i\geq j+2$, we can assume that for $k=a$, $ a\geq j+2$, the following statement is true.
\begin{align*}
\tilde{h}_{ij}å= &C_{i-1}[\prod_{m=1}^{i-(j+1)}(A_{i-m} + B_{i-m}L_{i-m})B_{j}L_{j} +
\sum_{c=1}^{i-(j+2)}[\prod_{m=1}^{c}(A_{i-m} + ... \\
&B_{i-m}L_{i-m})B_{i-(c+1)}L_{i-(c+1)}\prod_{b=c+2}^{i-j}(I - K_{i-b}C_{i-b})A_{i-b}] + ...\\
&B_{i-1}L_{i-1}\prod_{c=2}^{i-j}(I - K_{i-c}C_{i-c})A_{i-c}]K_{j}
\end{align*}
Then, for $k=a+1$, we can show that :
\begin{align*}
\tilde{h}_{(i+1)j}= &C_{i}[\prod_{m=1}^{i+1-(j+1)}(A_{i+1-m} + B_{i+1-m}L_{i+1-m})B_{j}L_{j} +...\\
&\sum_{c=1}^{i+1-(j+2)}[\prod_{m=1}^{c}(A_{i+1-m} + B_{i+1-m}L_{i+1-m})...\\
&B_{i+1-(c+1)}L_{i+1-(c+1)}\prod_{b=c+2}^{i+1-j}(I - K_{i+1-b}C_{i+1-b})A_{i+1-b}] +...\\
& B_{i+1-1}L_{i+1-1}\prod_{c=2}^{i+1-j}(I - K_{i+1-c}C_{i+1-c})A_{i+1-c}]K_{j}
\end{align*}
\section*{Appendix B: Defining matrices of simulation}
Defining matrices ($A_{i},B_{i},C_{i},i=1,2,3$) are as follows:
\begin{equation*}
 A_{1} = \begin{bmatrix}
 0.98451502& 0.10019498& 0.71348497\\
0.14298264& 0.6412398& 0.90647641 \\
0.58426722& 0.35536841& 0.47612775
\end{bmatrix}
\end{equation*}

\begin{equation*}
 B_{1} = \begin{bmatrix}
 0.44096615& 0.65555366& 0.94144979\\
0.78338986& 0.9915377& 0.04527771 \\
0.65264265& 0.71571167& 0.04051945
\end{bmatrix}
\end{equation*}

\begin{equation*}
 C_{1} = \begin{bmatrix}
 0.253187& 0.05120722& 0.11092476\\
0.29308483& 0.25376252& 0.27890331\\
0.75454911& 0.69534419& 0.84689801\\
0.67852479& 0.94239412& 0.47245499\\
0.45955921& 0.70151646& 0.8589794
\end{bmatrix}
\end{equation*}

\begin{equation*}
 A_{2} = \begin{bmatrix}
 0.32555806& 0.69552568& 0.4198415\\
0.15818161& 0.98608914& 0.17239575 \\
0.08682796& 0.46574264& 0.64864652
\end{bmatrix}
\end{equation*}

\begin{equation*}
 B_{2} = \begin{bmatrix}
 0.10907593& 0.92440577& 0.2639907\\
0.13940577& 0.0693751& 0.07336545 \\
0.91392411& 0.00977986& 0.70578249
\end{bmatrix}
\end{equation*}

\begin{equation*}
 C_{2} = \begin{bmatrix}
 0.11854015& 0.82173999& 0.36687075\\
0.53914991& 0.06616444& 0.0640871  \\
0.2704268& 0.98044219& 0.05198996\\
0.8653151& 0.23836825& 0.53458056\\
0.2535729& 0.24849771& 0.15870048
\end{bmatrix}
\end{equation*}

\begin{equation*}
 A_{3} = \begin{bmatrix}
 0.2825282& 0.03752622& 0.54049816\\
0.72578386& 0.68528011& 0.71830077 \\
0.64645617& 0.89273244& 0.543886
\end{bmatrix}
\end{equation*}

\begin{equation*}
 B_{3} = \begin{bmatrix}
 0.7053606& 0.15796312& 0.3572694\\
0.72379339& 0.16706866& 0.50119868 \\
0.66340254& 0.80151632& 0.24965837
\end{bmatrix}
\end{equation*}

\begin{equation*}
 C_{3} = \begin{bmatrix}
 0.43486379& 0.02126384& 0.69090388\\
0.09041975& 0.74105159& 0.35007977  \\
0.76560823& 0.96178511& 0.02544355\\
0.41486178& 0.55222053& 0.89840115\\
0.14928482& 0.54467456& 0.23947464
\end{bmatrix}
\end{equation*}

\end{document}